\documentclass[10pt,aps,showpacs,superscriptaddress,nofootinbib,twocolumn,preprintnumbers,floatfix,showkeys]{revtex4-1}
\usepackage{graphicx,amsmath,amssymb,soul,enumitem,bbm,dcolumn}
\usepackage{booktabs,multirow,array,slashed,dsfont,kantlipsum}
\usepackage[dvipsnames]{xcolor}
\usepackage{nicefrac}
\usepackage[utf8]{inputenc}
\usepackage[caption=false]{subfig}
\allowdisplaybreaks
\usepackage{hyperref}
\hypersetup{colorlinks=true,urlcolor=PineGreen,linkcolor=PineGreen,citecolor=PineGreen}
\begin{document}

\title{Finite-volume spectrum of $\pi^+\pi^+$ and $\pi^+\pi^+\pi^+$ systems}

\author{M.\ Mai}
\email{maximmai@gwu.edu}
\affiliation{Institute for Nuclear Studies and Department of Physics, The George Washington University, Washington, DC 20052, USA}
\author{M.\ D\"oring}
\email{doring@gwu.edu}
\affiliation{Institute for Nuclear Studies and Department of Physics, The George Washington University, Washington, DC 20052, USA}
\affiliation{Thomas Jefferson National Accelerator Facility, Newport News, VA 23606, USA}


\begin{abstract}
The ab-initio understanding of hadronic three-body systems above threshold, such as exotic resonances or the baryon spectrum, requires the mapping of the finite-volume eigenvalue spectrum, produced in lattice QCD  calculations, to the infinite volume. We present the first application of such a formalism to a physical system in form of three interacting positively charged pions. The results for the ground state energies agree with the available lattice QCD results by the NPLQCD collaboration at unphysical pion masses. Extrapolations to physical pion masses are performed using input from effective field theory.
The excited energy spectrum is predicted. This demonstrates the feasibility to determine three-body amplitudes above threshold from lattice QCD, including resonance properties of axial mesons, exotics, and excited baryons.
\end{abstract}



\maketitle

\paragraph*{\bf Introduction}
Many pressing questions in hadronic physics require the understanding of three-body systems above threshold. An example is the emblematic Roper resonance $N(1440)1/2^+$ that, despite its low mass, couples strongly to the $\pi\pi N$ channel leading to a very non-standard line shape and complicated analytic structure~\cite{Doring:2009yv}. This resonance is not only difficult to directly detect in experiment but also considerably lighter than the $N(1535)1/2^-$ (parity-partner puzzle), a phenomenon that is difficult to explain in the quark model~\cite{Loring:2001kx}. In general, almost all excited baryons have sizable couplings to $\pi\pi N$ states making the understanding of the three-body problem mandatory for the entire sector. Due to their key role in understanding confinement and other properties of Quantum Chromodynamics (QCD), and due to the {\it missing resonance problem}, excited baryons are subject of large experimental campaigns at Jefferson Lab, ELSA, MAMI and other facilities~\cite{Crede:2013sze, Aznauryan:2012ba, Klempt:2009pi}. 

Quantifying multi-neutron forces is also necessary for the equation of state of neutron matter in the extreme conditions of a neutron star~\cite{Baym:2017whm}. Recent advances in lattice QCD (LQCD) on few-nucleon systems~\cite{Beane:2012vq,Savage:2016egr} complement dedicated experimental programs, e.g., at the FRIB facility~\cite{Geesaman:2015fha}.

Three-body effects play also a crucial role in the understanding of axial mesons like the $a_1(1260)\to\pi\rho\to 3\pi$ and exotics whose existence would be a direct signal of gluon dynamics at low energies. The first claim for an exotic meson, the $J^{PC}=1^{-+}$ $\pi_1(1600)$ \cite{Alekseev:2009aa}, was made by the COMPASS collaboration by analyzing the three-pion final state; the Jefferson Lab Hall D GlueX experiment is designed to search for exotics for these and other produced mesons.

The calculation of the excited hadron spectrum and its properties, directly in terms of the fundamental degrees of freedom of QCD, has become possible in recent years~\cite{Briceno:2017max}; in LQCD, the QCD path integral is discretized and calculated numerically in a finite volume with periodic boundary conditions, leading to a discrete energy spectrum in contrast to the continuous spectral density of scattering states in the infinite volume. These finite-volume effects become even more relevant for quark masses coming closer to the physical ones as, e.g., bound states become resonances in certain systems~\cite{Guo:2018zss,Briceno:2016mjc}. Furthermore, the limit $L\to\infty$ does not provide direct access to the scattering amplitude at the Mandelstam $s+i\epsilon$, either, because this limit does not commute with the $\epsilon\to 0$ prescription (see e.g. ~Ref.~\cite{Agadjanov:2016mao}). However, as shown long ago by L\"uscher, each eigenvalue can be mapped to a phase-shift~\cite{Luscher:1986pf, Luscher:1990ux} in the elastic region of $2\to2$ scattering. On the other side, the $3\to 3$ reaction has eight independent kinematic variables while the $2\to 2$ reaction has only two; clearly, an entirely new formalism is necessary to map the eigenvalue spectrum to the physical amplitude.

To connect {\it ab-initio} LQCD to the rich resonance phenomenology, there has been much progress to find finite-volume formalisms for three-body systems above threshold~\cite{ Guo:2018ibd, Doring:2018xxx, Romero-Lopez:2018rcb, Klos:2018sen,  Mai:2017bge, Hammer:2017kms, Briceno:2017tce, Briceno:2018aml, Briceno:2018mlh, Hansen:2015zga, Hansen:2014eka,  Guo:2017ism, Polejaeva:2012ut, Briceno:2012rv, Kreuzer:2012sr}. 
For a recent review, see Ref.~\cite{Hansen:2019nir}.
Here, we apply, for the first time,  a  relativistic finite-volume formalism to a physical system. In particular, the formal developments of Refs.~\cite{Doring:2018xxx, Mai:2017bge} are used to analyze the $\pi^+\pi^+\pi^+$ system calculated by the NPLQCD collaboration~\cite{Beane:2007es,Detmold:2008fn}. Historically, the infinite-volume extrapolation of $2\pi^+$ system was the first physical application of the original L\"uscher formalism~\cite{Sharpe:1992pp, Kuramashi:1993ka, Gupta:1993rn, Yamazaki:2004qb, Aoki:2005uf}; similarly, the $3\pi^+$ system is the first system to make meaningful application of a three-body formalism for excited levels which is the subject of this Letter. 

The $3\pi^+$ system is, to a good approximation, free from problems of coupled channels and spin. With the applicability to real-world LQCD data demonstrated here, the pertinent extension of the formalism to analyze the Roper resonance and excited (exotic) mesons is the next milestone. Yet, three-body LQCD energy eigenvalues exist already~\cite{Lang:2014tia,Lang:2016hnn,Kiratidis:2016hda,Woss:2018irj} and such extensions are timely. For example, puzzling LQCD results on the Roper resonance~\cite{Lang:2016hnn} call for a better understanding of three-body effects in the coupled channels $\pi N,\,f_0(500)N,\,\pi\Delta,\,\rho N$, \dots.

Our program consists of prediction of the full finite-volume spectrum up to $4m_\pi$ using experimentally available data. Subsequently, we will fix the remaining parameter (genuine three-body coupling) to the ground-state energy level of the $\pi^+\pi^+\pi^+$ system~\cite{Beane:2007es,Detmold:2008fn}, predicting higher levels up to the $5\pi$ threshold. We note that in the long run, i.e., once excited levels are calculated in LQCD, the reversed procedure is the goal of this research direction. Namely, the extraction of the three-body amplitude from ab-initio Lattice QCD calculations of two- and three-body finite-volume spectra (see, e.g., Ref.~\cite{Briceno:2013bda} for such a procedure in the two-nucleon case).

\paragraph*{\bf Formalism}
In a cube of size $L$, periodic boundary conditions restrict the three momenta to values of ${\bf q}\in \frac{2\pi}{L}\,\mathds{Z}^3$. Consequently, the continuous scattering spectrum of the QCD Hamiltonian reduces to a tower of discrete energy eigenvalues and rotational symmetry is broken. For the mapping from finite to infinite volume, we implement these changes in a three-body amplitude to describe eigenvalues calculated in LQCD and then evaluate the same amplitude in the infinite volume~\cite{Mai:2017bge, Doring:2018xxx}. This also allows to simultaneously extract scattering amplitudes at different kinematic points which reduces the mentioned under-determination problem tied to the multiple kinematic variables in the three-body problem.

The relativistic amplitude fulfills three-body unitarity~\cite{Mai:2017vot} which has the advantage that one explicitly knows when all three particles are on-shell. Only these configurations lead to singularities of the scattering matrix in the finite volume and, thus, determine the leading power-law finite-volume effects. In contrast, off-shell configurations lead only to exponentially suppressed effects and are omitted like in the original L\"uscher approach for two particles. 

Without loss of generality, we choose a parametrization of the two-body sub-amplitudes by a tower of ``isobars'' which in the present case consists only of the isospin $I=2$ $\pi^+\pi^+$ S-wave scattering amplitude (without left-hand cuts).
The full three-body scattering amplitude can then be written in operator notation as $\mathsf{\hat T=v(\tau^{-1}+B)^{-1}v}$. Here $\mathsf{v}$, $\mathsf{\tau}$ and $\mathsf{B}$ denote the isobar dissociation vertex, isobar propagator and isobar-spectator interaction kernel, respectively. Three-body unitarity~\cite{Mai:2017vot} determines the imaginary parts of $B$ and $\tau$, such that, e.g., the former can be written as a one-pion-exchange diagram plus an unknown real-valued function $C$. Overall, this fully relativistic scattering equation is given by a set of three dimensional integral equation, which can be depicted by the following expansion,
\begin{center}
\includegraphics[width=0.99\linewidth]{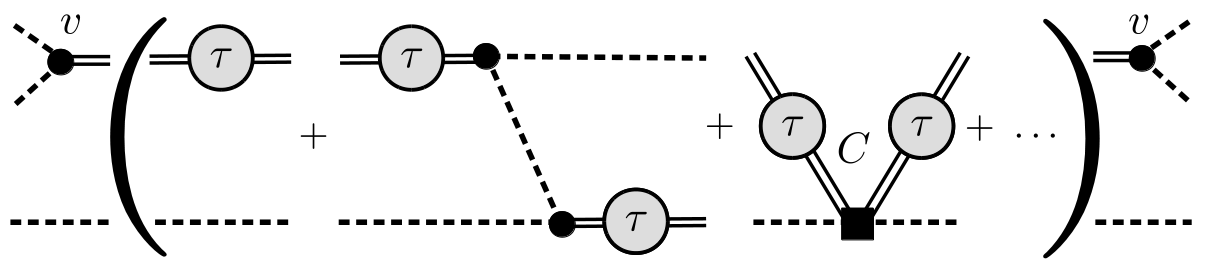}
\end{center}
where single dashed and double lines denote pions and isobar propagators, respectively. For the formal derivation of $\hat T$ see Refs.~\cite{Mai:2017bge, Mai:2017vot} as well as Ref.~\cite{Jackura:2018xnx} for a discussion of analytic properties of such an amplitude. Here, we note that the formalism is a priori dispersive and not tied to an expansion in Feynman diagrams.

Through the discretization of momenta in finite volume, the above equation for the scattering amplitude becomes a real-valued matrix-equation, whose singularities correspond to the eigenvalues of the QCD Hamiltonian. In a formal language, the three-body eigenvalues are determined through the quantization condition~\cite{Doring:2018xxx}
\begin{align}
\label{eq:QC3}
\det\left(B^{\Gamma ss'}_{uu'} +\frac{2E_s \,L^3}{\vartheta(s)}
\tau_s^{-1}\delta_{ss'}\delta_{uu'}\right)=0\,,
\end{align}
where the determinant is taken with respect to the basis index $u{}^{(}{'}{}^{)}$ for a given irrep $\Gamma\in\{A_1,A_2,E,T_1,T_2\}$ and $s{}^{(}{'}{}^{)}$ denotes the sets of momenta related by cubic symmetry (``shells'') with cardinality $\vartheta(s)$. In the following, we work in the center of mass system of three pions with the total four-momentum $P=(W_3,\boldsymbol{0})$ and pion energies denoted by $E_s:=E_{\bf p}:=\sqrt{m_\pi^2+{\bf p}^2}$. Finally, dealing with a three-pion system of maximal isospin in S-wave, only one isobar ($\pi^+\pi^+$ sub-system) is of interest, while the irrep will be fixed throughout this work to $\Gamma=A^+_1$.

\begin{figure*}[thb]
\includegraphics[width=0.49\linewidth]{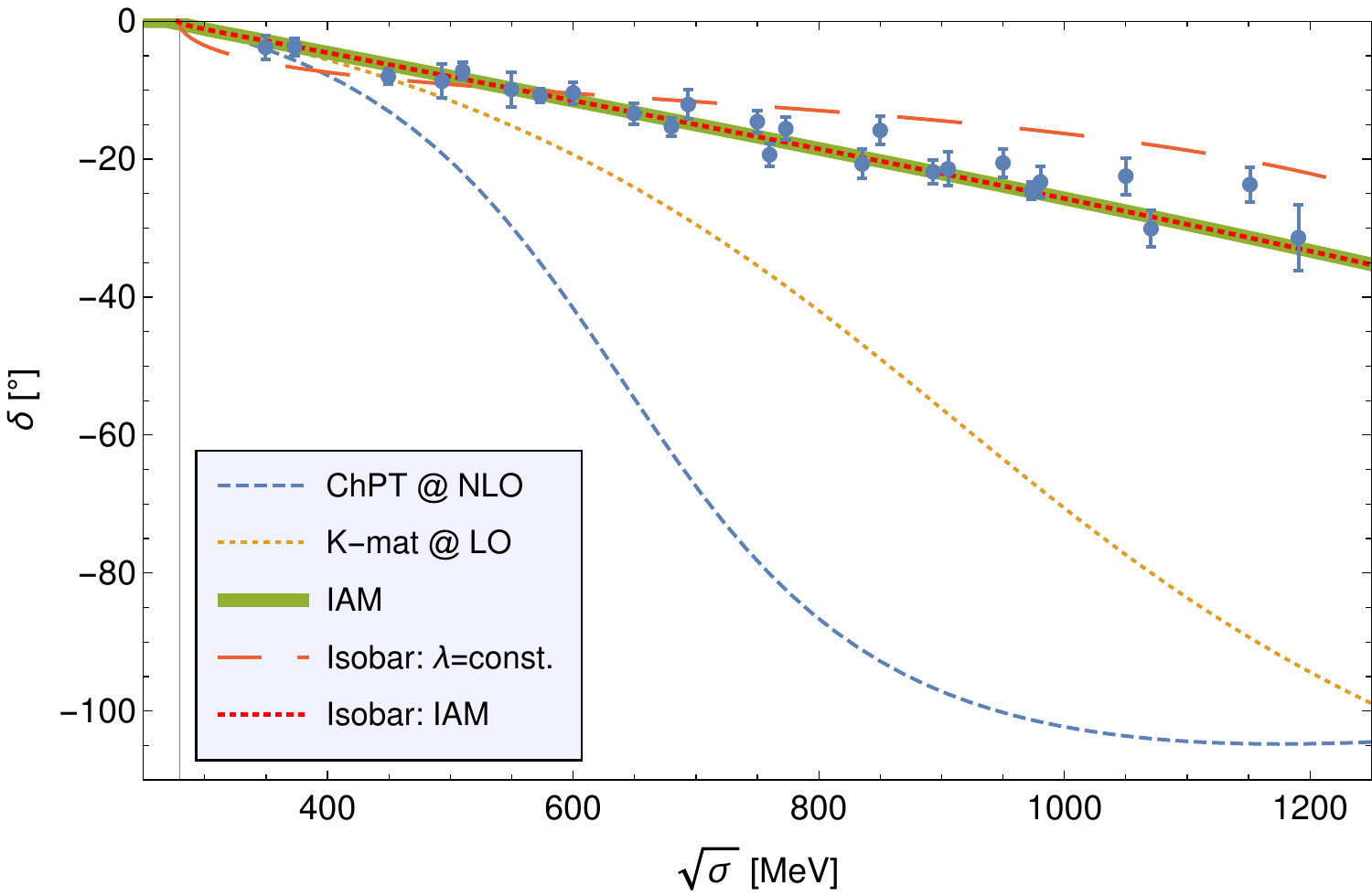}
  ~
  \includegraphics[width=0.49\linewidth]{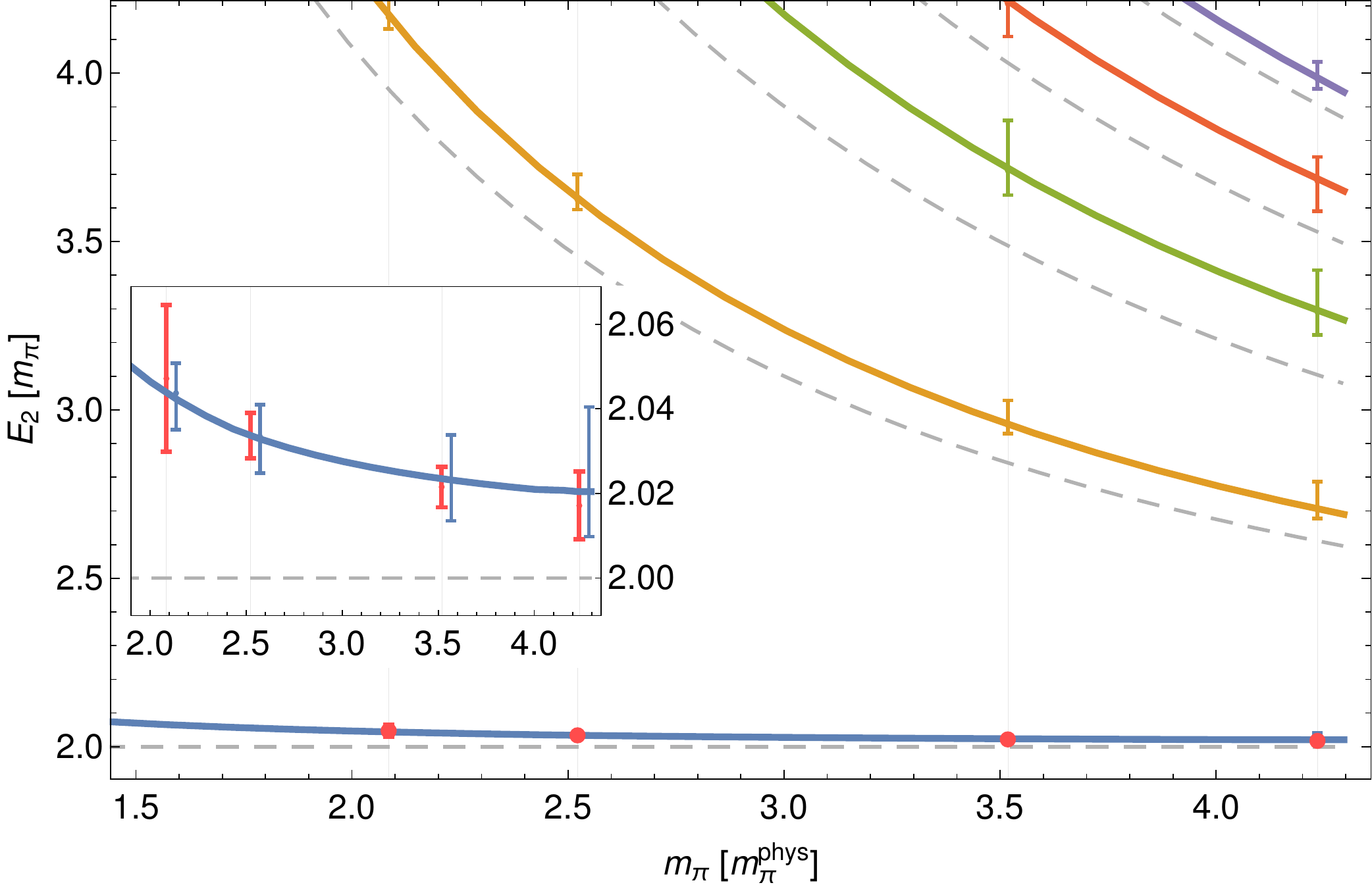}
\caption{
Left: Comparison of available phase-shift data~\cite{Rosselet:1976pu,Janssen:1994wn,Estabrooks:1974vu} with the prediction of the considered models.
Right: Prediction of two-body energy levels (full) as a function of $m_\pi$ with dashed lines denoting non-interacting levels and error bars indicating the prediction uncertainty (see Tab.~\ref{tab:EEV} in Appendix~\ref{app:numres}). The inset shows a zoom-in on the ground level, where the red points denote the result of the lattice calculation~\cite{Beane:2007es,Detmold:2008fn}.}
\label{pic:2body}
\end{figure*}

The projection techniques to the irreps of the cubic group can be found in Ref.~\cite{Doring:2018xxx}, while  the terms $B$ and $\tau$ are given  for convenience in Appendix~\ref{app:details}. The dissociation vertex is given by ${v=f(Q^2)\lambda(\sigma)}$, where $Q$ and $\sigma$ denote the four-momentum difference of the isobar decay-products and the invariant mass squared of the isobar, respectively. The form factor $f(Q^2)$ yields a smooth cutoff of an otherwise log-divergent self-energy part of the isobar propagator. Thus, individual kernels entering the quantization condition~\eqref{eq:QC3} are only defined in a given regularization scheme. Ultimately, this dependence cancels out after renormalizing (e.g., $C$) as described below. Specifically, we chose here ${f(Q^2)=1/(1+e^{-(\Lambda/2-1)^2+ Q^2/4})}$ in units of the pion mass.

\paragraph*{\bf Results: Two-body Subsystem} 
The three-body scattering amplitude~\cite{Mai:2017vot} corresponding to the quantization condition~\eqref{eq:QC3} fulfills two- (in every sub-channel) and three-body unitarity by construction. The corresponding normalized two-body scattering amplitude, projected to S-wave, reads in operator language $\mathsf{\hat T_2=v\hat \tau v}$ for
\begin{align}
\label{eq:T2}
\frac{\hat\tau(\sigma)^{-1}}{32\pi\lambda(\sigma)^2}
=&K^{-1}(\sigma)\\\nonumber
& 
-\sum_{\pm}\int \frac{d^3\boldsymbol{k}}{(2\pi)^3} 
\frac{(f((\sqrt{\sigma}\pm 2E_k)^2-4\boldsymbol{k}^2))^2}
{4E_{\boldsymbol k}\sqrt{\sigma}(\sqrt{\sigma}\pm2E_{\boldsymbol k})} \,, 
\end{align}
where $\hat\tau$ is the infinite-volume counterpart of the isobar propagator $\tau$ and $K(\sigma)=\lambda(\sigma)^2/(\sigma-M_0^2)$. The yet unknown parameters ($\lambda$ and $M_0$) will be constrained using the available experimental phase-shifts~\cite{Rosselet:1976pu,Janssen:1994wn,Estabrooks:1974vu} in the following.

We have explored several ansatzes for the functional form of $\lambda$, collecting the outcome as depicted in Fig.~\ref{pic:2body}. In the simplest case ($\lambda={\rm const.}$) we fit $\lambda$ and $M_0$ to the experimental data obtaining only fair agreement with data; also, no meaningful chiral extrapolation can be provided. The perturbative amplitude of the next-to-leading chiral order~\cite{Gasser:1983yg} and the unitarized amplitude using only the leading chiral order describe the data well only in a close proximity to the $\pi\pi$-threshold. We found that the Inverse Amplitude Method (IAM), see Refs.~\cite{Truong:1988zp,Nebreda:2010wv}, i.e. $T_{\rm LO}^2/(T_{\rm LO}-T_{\rm NLO})$, shows the best agreement with the data. Furthermore, it can be expressed in the form of Eq.~\eqref{eq:T2}, demanding
\begin{align}
\lambda^2=(M_0^2-\sigma)\left(\frac{d}{4\pi^2}+\frac{T_{\rm LO}-\bar T_{\rm NLO}}{T_{\rm LO}^2}\right)^{-1}\,,
\end{align}
where $\bar T_{\rm NLO}$ denotes the next-to-leading order chiral amplitude~\cite{Gasser:1983yg} without the s-channel loop, which depends on low-energy constants (LECs) taken from the same reference. The constant $d$ compensates for the fact that dimensional regularization was used in Ref.~\cite{Gasser:1983yg}, while in the present ansatz we use form-factors to regulate the divergences. We found that choosing $\Lambda=42$  corresponds to ${d=0.86}$ such that the both formulations of the scattering amplitude coincide perfectly (see Fig.~\ref{pic:2body}) which also holds for all pion masses in question. The scattering lengths read for unphysical pion masses
\begin{align}
\label{eq:scatteringlenghts}
&a_{291}=-0.1478_{-0.0550}^{+0.0356}\,,\quad
a_{352}=-0.2016_{-0.1008}^{+0.0663}\,,\\\nonumber
&a_{491}=-0.3622_{-0.1395}^{+0.1914}\,,\quad
a_{591}=-0.5406_{-0.1728}^{+0.3645}\,,
\end{align}
which agree with previous LQCD results, e.~g., Ref.~\cite{Dudek:2010ew}, while  $a_{139.57}=-0.0433(37)$ for the physical one compares perfectly with $-0.0444(10)$ from the Roy equation analysis of Ref.~\cite{Ananthanarayan:2000ht}. The uncertainties are determined from sampling of the LECs taking uncorrelated error bars from Ref.~\cite{Gasser:1983yg}. Note that the regulator dependence in the form factor plays no role; all we have done here is parametrizing the physical region of $I=2$ $\pi\pi$ scattering; if we had to change the regularization we simply needed to renormalize by re-fiting the LECs to the $\pi\pi$ phase-shifts. 

Unitarity, correct description of data and proper chiral behavior are the only features required for the realistic prediction of the finite-volume spectrum. Therefore, having fixed $\lambda$ as described before, we predict the $\pi^+\pi^+$ finite-volume spectrum ($L=2.5$~fm), determining the roots of $\tau^{-1}$ in the two-body energy $\sqrt{\sigma}$. Note that this is equivalent to L\"uscher's method~\cite{Luscher:1986pf, Luscher:1990ux} up to exponentially suppressed terms~\cite{Doring:2011vk}.

The result is depicted in the right panel of Fig.~\eqref{pic:2body}, while the numerical values for the physical and the pion masses used in the lattice calculation~\cite{Beane:2007es,Detmold:2008fn} are collected in Tab.~1 in Appendix~\ref{app:numres}. The quoted error bars are determined in a 40-point sampling varying the LECs from~\cite{Gasser:1983yg}. 

The result for the post-dicted ground state level agrees nicely with the lattice calculation ($\chi_{\rm p.p.}^2=0.35$), and is in agreement with the large-volume expansion formula~\cite{Detmold:2008gh} using the scattering lengths from Eq.~(\ref{eq:scatteringlenghts}) as input. Notably and unexpectedly, the IAM-like chiral extrapolation seem to work well up to very high pion masses. The excited energy levels shown in the right panel of Fig.~\ref{pic:2body} are predictions. Note that no 4-particle cuts have been discussed, such that the prediction is quoted up to ${\sqrt{\sigma}=4m_\pi}$.

\paragraph*{\bf Results: Three-body Energy Shift}
With two-body input at hand showing good agreement with the LQCD data, we turn now to the main point of the present paper, the finite-volume spectrum of the $\pi^+\pi^+\pi^+$ system. In a three-particle system, the invariant mass of the two-particle system can be sub-threshold ($\sqrt{\sigma_{\boldsymbol{q}}}<2m_\pi$) for a sufficiently large momentum of the spectator $\boldsymbol{q}$. Note that only right-hand (physical) two-body singularities are included in the derivation of the three-body scattering amplitude~\cite{Mai:2017bge,Mai:2017vot}, leading to  the quantization condition~\eqref{eq:QC3}. Furthermore, in the absence of two-body bound states, the infinite-volume two-body amplitude, derived through a dispersion relation~\cite{Mai:2017vot}, has to be real and regular in the sub-threshold region. In the three-body framework, this two-body sub-threshold contribution is compensated by the (still) unknown real function $C$. Furthermore, in finite volume, corrections from this region are exponentially suppressed. In summary, at some $\sigma_0$ in the unphysical region one can simply set $K_{\boldsymbol{q}}^{-1}$ to a (real) constant that is smoothly connected to the physical region, where it reproduces the IAM-type of scattering amplitude as described before. Below, we check the dependence on $\sigma_0$ explicitly.

The remaining unknown piece of the quantization condition~\eqref{eq:QC3} is the 3-body interaction term $C$, which can only be determined from a fit to data. Fortunately, lattice data are available for the ground state~\cite{Beane:2007es,Detmold:2008fn} in the same setup as for the two-pion system. Note that in general, $C$ is a function of the in/outgoing spectator momenta ($\boldsymbol{q}/\boldsymbol{p}$), total energy $W_3$, and $m_\pi$. We found that already the simplest choice $C_{\boldsymbol{q}\boldsymbol{p}}=c\,\delta^{(3)}(\boldsymbol{p}-\boldsymbol{q})$ leads to a good fit to the ground-state energies $E_3^1$~\cite{Beane:2007es,Detmold:2008fn}, i.e. $\chi^2_{\rm dof}=0.05$ for $c_{\rm fit}=\left(0.2\pm 1.5\right)\cdot 10^{-10}$, i.e., a value compatible with zero. The statistical and systematic data uncertainty were added for this fit, which explains the low value of $\chi^2_{\rm dof}$. 

\begin{figure}[t]
\includegraphics[width=0.99\linewidth]{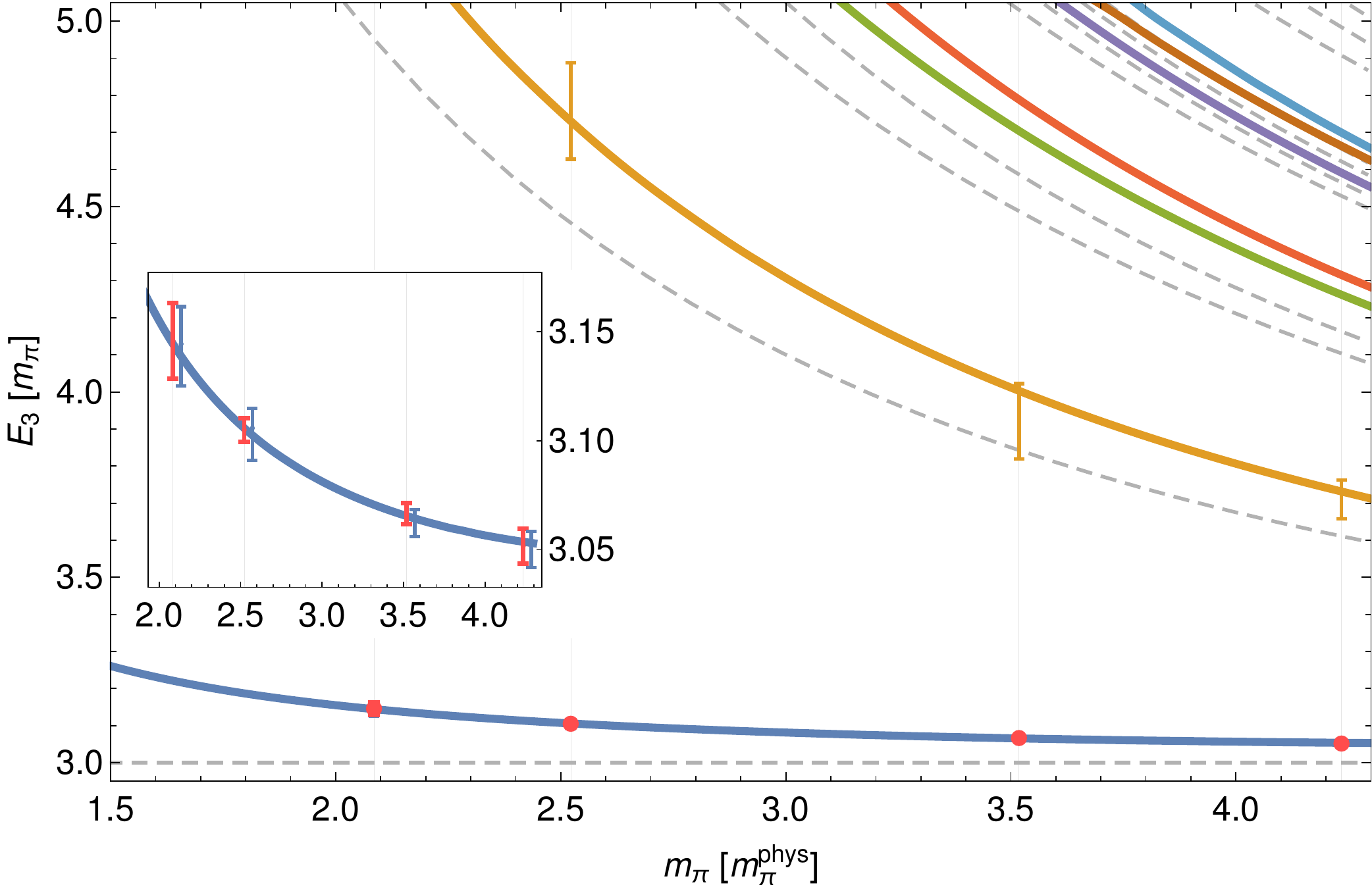}
\caption{
Prediction of excited energy levels for the $\pi^+\pi^+\pi^+$-system as a function of pion mass with non-interacting levels represented by dashed lines. The uncertainty on the first two levels is indicated. The inset shows the zoom-in on the ground level, where the lattice data~\cite{Beane:2007es,Detmold:2008fn} are shown in red.
Note that $m_\pi L<4$ for the lowest pion mass of the data (see text).}
\label{pic:E3}
\end{figure}

The result of the fit to the ground level as well as prediction of higher levels are depicted in Fig.~\ref{pic:E3} with the uncertainties from a sampling of LECs as before. We observe, that while the first two levels have rather small error bars, higher levels appear in two dense clusters (3,4) and (5,6,7) and overlap withing the 1$\sigma$ uncertainty bands. For clarity of the presentation we do not show the latter in the Fig.~\ref{pic:E3} but Tab.~\ref{tab:EEV} in Appendix~\ref{app:numres} quotes all uncertainties. This shows that when the levels are obtained from a lattice calculation, they can put new strong bound on the two-body amplitudes, which are the main source of uncertainty for the calculated three-body energy eigenvalues. Note that for $m_\pi<315$~MeV (i.e., for the lowest pion mass of the NPLQCD calculation) there might be larger exponentially suppressed finite-volume according to the rule of thumb that such effects can be safely neglected only for $m_\pi L>4$.

As an additional check we have fitted the ground state levels $E_3^1$~\cite{Beane:2007es,Detmold:2008fn} using the large-volume expansion formula~\cite{Detmold:2008gh} using our scattering lengths~\eqref{eq:scatteringlenghts} and adjusting the unknown three-body contribution $\eta_3^L$ (Eqs.~(1-5) of~\cite{Detmold:2008gh}). The fit yields $\chi^2_{\rm dof}=1.32$ for $\eta_3^L=1.8\cdot 10^{-12}$ and 
$E_3^1=\{3.1277, 3.1003,3.0695,3.0623\}~m_\pi$ for $m_\pi=\{291,352,491,591\}$~MeV, respectively. It is interesting to see that not only this confirms our result  for the lowest level, but also that the genuine three-body contribution is similar to $c$ determined before, keeping in mind that the latter was introduced on the level of amplitudes and not a Hamiltonian as $\eta_3^L$.

On a qualitative level, we observe that the energy levels shown in Fig.~\ref{pic:E3} mimic the pattern of the non-interacting ones shifted to higher energies. This is similar to the two-body case with the novelty that interacting energy levels do not always occur between two non-interacting ones.

Our method relies on regularization and renormalization through the three-body interaction in intermediate steps. We therefore discuss the renormalization procedure and independence of the results on the regulator. Two different cutoffs have been introduced in our two- and three-body calculation: (a) A form-factor $f$ regulates the log-divergent integral in the self-energy~\eqref{eq:T2} and also the  divergence of the three-body equation, i.e., as $s,s'\to\infty$ in the quantization condition~\eqref{eq:QC3}; (b) The determinant in \eqref{eq:QC3} is taken over a finite number of shells, which effectively introduces a hard (spectator) momentum cutoff. Regarding (a), we have checked that the eigenvalue spectrum in the two- and three-body case barely changes if one changes the form factor from $f(Q^2)=1/(1+e^{-(\Lambda/2-1)^2+ Q^2/4})$ to $f(Q^2)=\Lambda^4/(\Lambda^4+Q^4)$.
These changes vanish  when the parameters of the two- and three-body amplitudes (LECs and three-body force $C$ in the present case) are renormalized accordingly (see also the remark after Eq.~\ref{eq:scatteringlenghts} on the renormalization of the two-body amplitude). To demonstrate such a process we turn now to (b), which occurs due to a truncation of the matrix in the argument of Eq.~\eqref{eq:QC3} with respect to the number of shells $s,s'\leq s_{\rm max}$. Throughout the work we have considered 20 shells, which corresponds to a momentum cutoff of $2.1$~GeV. Truncating the matrices at a lower number of shells without changing the three-body force, the eigenvalue spectrum changes as depicted in Appendix~\ref{app:regulator}. However, renormalizing the three-body force ${c_{\rm fit}=0.2\cdot 10^{-10}}$ as
\begin{center}
\renewcommand\arraystretch{1.2}
\begin{tabular}{|c|c|c|c|}
\hline
$s_{\rm max}$        & 2 ($\sim 0.50$ GeV)     & 4 ($\sim 0.86$ GeV)     &20 ($\sim 2.1$ GeV)\\
\hline
$c/c_{\rm fit}$ & 0.97  & 0.99  &1.00 \\
\hline
\end{tabular}
\end{center}
\noindent
returns the original result for all ground state levels $E_3^1$.
In summary, we have demonstrated the renormalization with a constant three-body term $C$. With future data on excited levels from LQCD it will be possible to learn more about the energy and momentum dependence of the three-body force.

\paragraph*{\bf Conclusion}
The finite volume spectrum for the $\pi^+\pi^+$ and $\pi^+\pi^+\pi^+$ systems has been analyzed using a finite-volume method based on three-body unitarity that identifies all power-law finite-volume effects of a three-body system. Using experimental data and a non-perturbative ansatz for the two-body amplitude we have predicted the $\pi^+\pi^+$ energy levels in finite volume which are in perfect agreement with the lattice data available for the ground state. Finally, using this input and fitting the genuine three-body contact term to the threshold level determined by the NPLQCD collaboration we have predicted the excited level spectrum of the $\pi^+\pi^+\pi^+$ system up to $W_3=5\,m_\pi$. This is the first prediction of excited levels in a physical three-body system. Possible sources for systematics (choice of parametrization of the two-body amplitude and its sub-threshold behavior, three-body force, and regularization) and statistical uncertainties have been identified and estimated.

In summary, we have demonstrated how the lattice results for systems with three hadrons can be connected with the experimental data. While the extensions to coupled-channels, $2\to 3$ processes, unequal masses and isobars with spin are work in progress, this pioneering study opens the way for analysis of hadronic systems like the Roper resonance in the future.

\bigskip

\begin{acknowledgments}
We thank C.~Urbach, A.~Rusetsky, A.~Alexandru and R.~Briceño for inspiring discussions and E.~Sismanidou for patience and backing. The work of MM was supported by the German Research Association (MA 7156/1) and by the National Science Foundation grant no. PHY-1452055. MD acknowledges support by the National Science Foundation (grant no. PHY-1452055) and by the U.S. Department of Energy, Office of Science, Office of Nuclear Physics under contract no. DE-AC05-06OR23177 and grant no. DE-SC0016582.
\end{acknowledgments}

\bibliography{MM-all-ref.bib}

\begin{widetext}
\newpage
\appendix

\begin{center}
\textbf{-- Supplemental Material --}
\end{center}
\section{Details on Formalism}
\label{app:details}

Explicit details on the formalism are provided that are not essential for the understanding of the main text. As explained there, the building blocks of the three-body formalism in finite volume are the isobar-spectator interaction kernel $B$ and the isobar propagator $\tau$ that enter the integral equation for the isobar-spectator interaction, ${T_{\bf qp}=B_{\bf qp}+\sum_{{\bf r}\in \frac{2\pi}{L}\,\mathds{Z}^3}B_{{\bf q}{\bf r}}\,\tau_{\bf r}\,T_{{{\bf r}{\bf p}}}}$, which leads to the quantization shown in Eq.~(1) in the main text. For complete details and derivations see Ref.~\cite{Mai:2017bge}. Before projection to the $A_1^+$ irrep~\cite{Doring:2018xxx}, $B$ and $\tau$ read
\begin{align}
\label{eq:btau}
B_{\boldsymbol{q}\boldsymbol{p}}
&=-
\frac{f((P-2q-p)^2)f((P-q-2p)^2)}
{2E_{\boldsymbol q+\boldsymbol{p}}
(W_3-E_{\boldsymbol{p}}-E_{\boldsymbol{q}}-E_{\boldsymbol{q}+\boldsymbol{p}})}
+C_{\boldsymbol{q}\boldsymbol{p}}\,,
\\
\tau^{-1}_{\boldsymbol{q}}
&=K^{-1}_{\boldsymbol{q}}
-\frac{J_{\boldsymbol{q}}}{L^3}
\sum_{\boldsymbol{x}\in \frac{2\pi}{L}\,\mathds{Z}^3}
\sum_\pm
\frac{
\left(
f\left(
   \left(
      P^*_{\boldsymbol{q}}\pm2k^*_{\boldsymbol{x},\boldsymbol{q}}
   \right)^2
 \right)
\right)^2}
{4 \sqrt{\sigma_{\boldsymbol{q}}}
E_{\boldsymbol{k}^*_{\boldsymbol{x},\boldsymbol{q}}}
\left(
\sqrt{\sigma_{\boldsymbol{q}}}\pm
2E_{\boldsymbol{k}^*_{\boldsymbol{x},\boldsymbol{q}}}
\right)}\,,\nonumber
\end{align}
where $\boldsymbol{p}/\boldsymbol{q}$ are the three-momenta of the in/outgoing spectator pions (being identical for the propagator) and $P^*_{\boldsymbol{q}}:=(\sqrt{\sigma_{\boldsymbol{q}}},\boldsymbol{0})$ is the four-momentum of the isobar (two-pion system) boosted to its reference frame with squared invariant mass $\sigma_{\boldsymbol{q}}=W_3^2+m_\pi^2-2W_3E_{\boldsymbol{q}}$. The three-momentum of pions $\boldsymbol{x}$ boosted by $\boldsymbol{q}$ is denoted by $\boldsymbol{k}^*_{\boldsymbol{x}\boldsymbol{q}}$ with $J_{\boldsymbol{q}}$ being the corresponding Jacobian. Furthermore, the form-factor $f(Q^2)$ yields a smooth cutoff of an otherwise log-divergent self-energy part of the isobar propagator (second term in $\tau^{-1}$ of Eq.~\eqref{eq:btau}). Note that this cutoff-dependence cancels in the full quantization condition, shown in Eq.~(1) in the main text, by the functions $C$ and $K$. None of the final results depend on this choice as discussed in the main text.


\section{Numerical Results}
\label{app:numres}

For completeness we provide the values for the two- and three-body finite-volume eigenvalues in Tab.~\ref{tab:EEV}. The values $E_2^i$ of the two-body system are shown in Fig.~1 (right panel) of the main text. The values $E_3^i$ of the three-body system are shown in Fig.~2 of the main text. Note that for $E_3^i,\, i\geq 3$ the error bars can overlap.
\renewcommand\arraystretch{1.3}
\begin{table*}[b!]
\begin{tabular}{|c|lllll|}
\hline
$~~~~~~~~~m_\pi$~[MeV]
&$~~~~139.57$ 
&$~~~~291$ 	
&$~~~~352$	 
&$~~~~491$		
&$~~~~591$ \\
\hline
\hline
$E_2^1~[m_\pi]~~~~~~~~~~~~$
&$~~~~2.1228_{-0.0069}^{+0.0068}$ 
&$~~~~2.0437_{-0.0086}^{+0.0071}$ 
&$~~~~2.0334_{-0.0086}^{+0.0076}$ 
&$~~~~2.0233_{-0.0098}^{+0.0105}$	
&$~~~~2.0204_{-0.0106}^{+0.0200}$ \\
~~~~~~~~~~~~Refs.~\cite{Beane:2007es,Detmold:2008fn}
&$~~~~-$
&$~~~~\mathbf{2.0471(27)(65)}$ 
&$~~~~\mathbf{2.0336(22)(22)}$ 
&$~~~~\mathbf{2.0215(16)(13)}$
&$~~~~\mathbf{2.0171(16)(19)}$\\
\hline
$E_2^2~[m_\pi]~~~~~~~~~~~~$
&$~~~~-$ 
&$~~~~-$ 
&$~~~~3.6245_{-0.0299}^{+0.0746}$ 
&$~~~~2.9556_{-0.0263}^{+0.0728}$	
&$~~~~2.7045_{-0.0271}^{+0.0827}$ \\
\hline
$E_2^3~[m_\pi]~~~~~~~~~~~~$
&$~~~~-$ 
&$~~~~-$ 
&$~~~~-$ 
&$~~~~3.7114_{-0.0737}^{+0.1482}$
&$~~~~3.2911_{-0.0688}^{+0.1241}$ \\
\hline
$E_2^4~[m_\pi]~~~~~~~~~~~~$
&$~~~~-$ 
&$~~~~-$ 
&$~~~~-$ 
&$~~~~-$	
&$~~~~3.6802_{-0.0902}^{+0.0707}$ \\
\hline
$E_2^5~[m_\pi]~~~~~~~~~~~~$
&$~~~~-$ 
&$~~~~-$ 
&$~~~~-$ 
&$~~~~-$	
&$~~~~3.9829_{-0.0299}^{+0.0500}$ \\
\hline
\multicolumn{6}{c}{~}\\[-10pt]
\hline
$E_3^1~[m_\pi]~~~~~~~~~~~~$ 
&$~~~~{3.6564}_{-0.0847}^{+0.1014}$
&$~~~~^{*}{}{3.1444}_{-0.0192}^{+0.0171}$
&$~~~~^{*}{}{3.1058}_{-0.0147}^{+0.0091}$
&$~~~~^{*}{}{3.0655}_{-0.0095}^{+0.0029} $
&$~~~~^{*}{}{3.0537}_{-0.0119}^{+0.0048} $\\
~~~~~~~~~~~~Refs.~\cite{Beane:2007es,Detmold:2008fn}
&$~~~~-$
&$~~~~\mathbf{3.1458(49)(125)}$ 
&$~~~~\mathbf{3.1050(27)(27)}$ 
&$~~~~\mathbf{3.0665(26)(22)}$ 
&$~~~~\mathbf{3.0516(27)(53)}$\\
\hline
$E_3^2~[m_\pi]~~~~~~~~~~~~$ 
&$~~~~-$
&$~~~~-$
&$~~~~{4.7301}_{-0.1027}^{+0.1577}$
&$~~~~{4.0031}_{-0.1836}^{+0.0196}$
&$~~~~{3.7315}_{-0.0742}^{+0.0309} $\\
\hline
$E_3^3~[m_\pi]~~~~~~~~~~~~$ 
&$~~~~-$
&$~~~~-$
&$~~~~-$
&$~~~~{4.7043}_{-0.5923}^{+0.0126} $
&$~~~~{4.2621}_{-0.1739}^{+0.0001} $\\
\hline
$E_3^4~[m_\pi]~~~~~~~~~~~~$ 
&$~~~~- $
&$~~~~-$
&$~~~~-$
&$~~~~{4.7890}_{-0.1722}^{+0.0506}$
&$~~~~{4.3155}_{-0.1341}^{+0.0837} $\\
\hline
$E_3^5~[m_\pi]~~~~~~~~~~~~$ 
&$~~~~-$
&$~~~~-$
&$~~~~-$
&$~~~~-$
&$~~~~{4.5913}_{-0.1995}^{+0.0001} $\\
\hline
$E_3^6~[m_\pi]~~~~~~~~~~~~$ 
&$~~~~-$
&$~~~~-$
&$~~~~-$
&$~~~~-$
&$~~~~{4.6634}_{-0.1070}^{+0.0001} $\\
\hline
$E_3^7~[m_\pi]~~~~~~~~~~~~$ 
&$~~~~-$
&$~~~~-$
&$~~~~-$
&$~~~~-$
&$~~~~{4.6995}_{-0.0661}^{+0.0001} $\\
\hline
\end{tabular}
\caption{
Energy levels of the $\pi^+\pi^+$ ($E_2$) and $\pi^+\pi^+\pi^+$ ($E_3$) system (level order is denoted by the superscript) with error bars from a sampling procedure described in the main text. The Lattice QCD results~\cite{Beane:2007es,Detmold:2008fn} are quoted in bold font with statistical and systematic uncertainties in first and second parenthesis, respectively. Eigenvalues fitted to the lattice data via the  three-body coupling $c$ are denoted by an asterisk while all $E_3^i,\,i>1$ are predictions.}
\label{tab:EEV}
\end{table*}

\section{Regulator dependence}
\label{app:regulator}

As explained in the main part of the paper, the number of shells is truncated when solving Eq.~(1) for LQCD energy eigenvalues. Effectively, this yields a "hard" spectator-momentum cutoff. Results of the paper have been obtained using 20 shells, which corresponds to the largest momentum of $2\pi/L\sqrt{18}\approx2.1$~GeV. Fig.~\ref{pic:shells} shows the determinant in Eq.~(1) as a function of the total energy $W_3$ at a representative pion mass of $m_\pi=400$~MeV and for shells $\{2, 4, 6, 8, 10, 12\}$ corresponding to momentum cutoffs of $\{0.496, 0.859, 1.109, 1.403, 1.488, 1.645\}$~GeV, respectively. The lattice size is chosen to be $L=2.5$~fm.

The dependence on the spectator momentum cutoff is very weak. Furthermore, for this figure we fixed the the three-body force $c$ to the central fitted value, $c=c_{\rm fit}=0.2\cdot 10^{-10}$. However, one should keep in mind that via renormalization the three-body force itself depends on the cutoff. In particular, for a each fixed momentum cutoff $c$ has to be re-fitted. In the main part of the manuscript this is explicitly demonstrated.

\section{Volume dependence}

Lattice calculations of systems with three hadrons require a considerable computational effort. Thus, it is of interest to have a prediction of the energy eigenvalues to guide such calculations at the first place. In Fig.~\ref{pic:Ldep} we present the volume dependence of the excited level spectrum at a representative pion mass of $m_\pi=291$~MeV. An interesting observation is that some levels (5 and 6) are very close to each other. This requires considerable resolution ($\mathcal{O}(1$~MeV)) of future lattice calculation if these levels need to be resolved.

\begin{figure*}[t!]
\begin{minipage}[t]{0.47\columnwidth}
\includegraphics[width=0.965\linewidth]{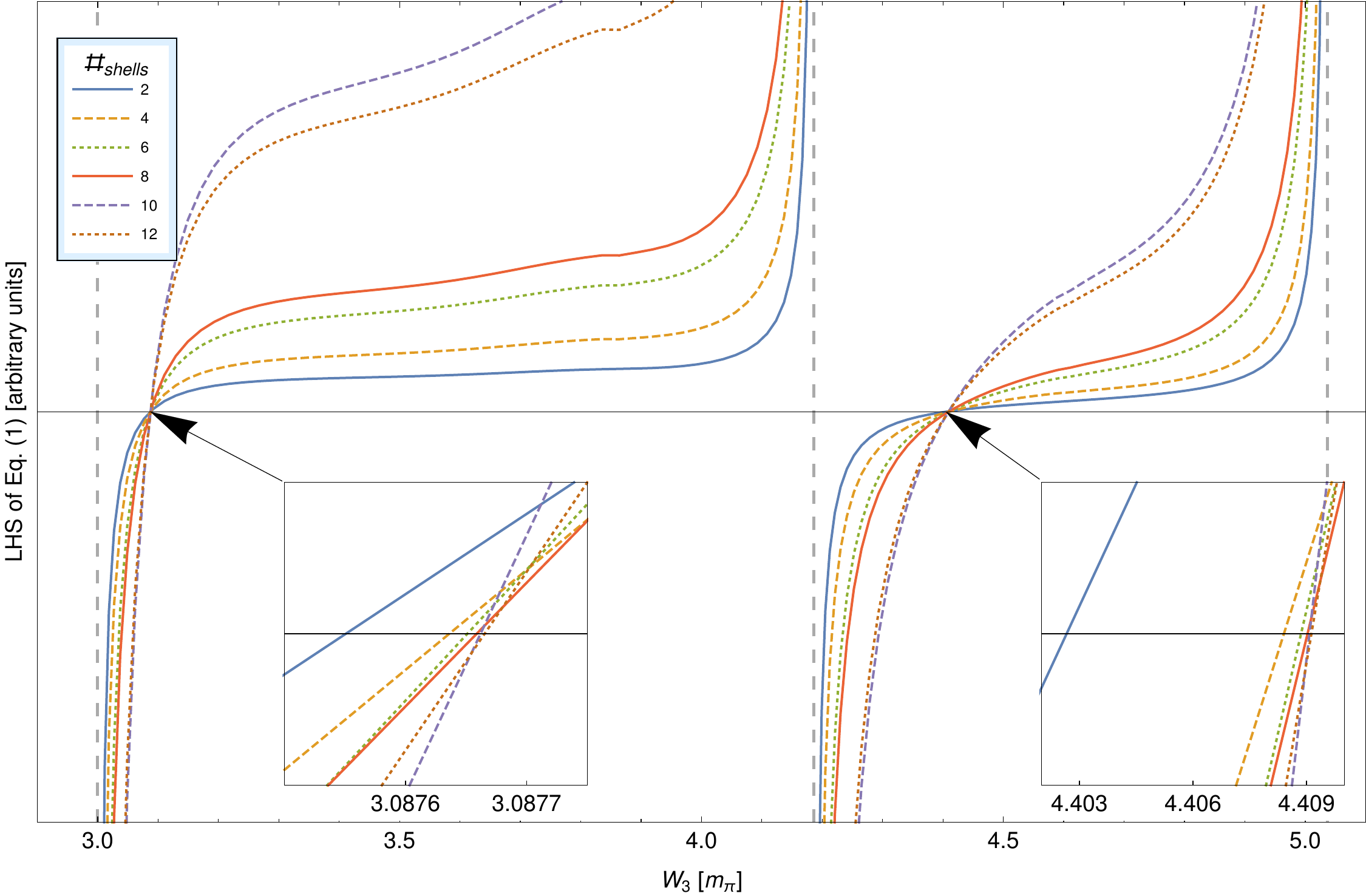}
\caption{
Momentum cutoff dependence of the three-body energy eigenvalues that are given by the zeros of the left-hand side (LHS) of Eq.~(1). Results are shown for different cutoffs as denoted in the legend. The dashed vertical lines denote the positions of the non-interacting levels.}
\label{pic:shells}
\end{minipage}
~~
\begin{minipage}[t]{0.47\columnwidth}
\includegraphics[width=0.99\linewidth]{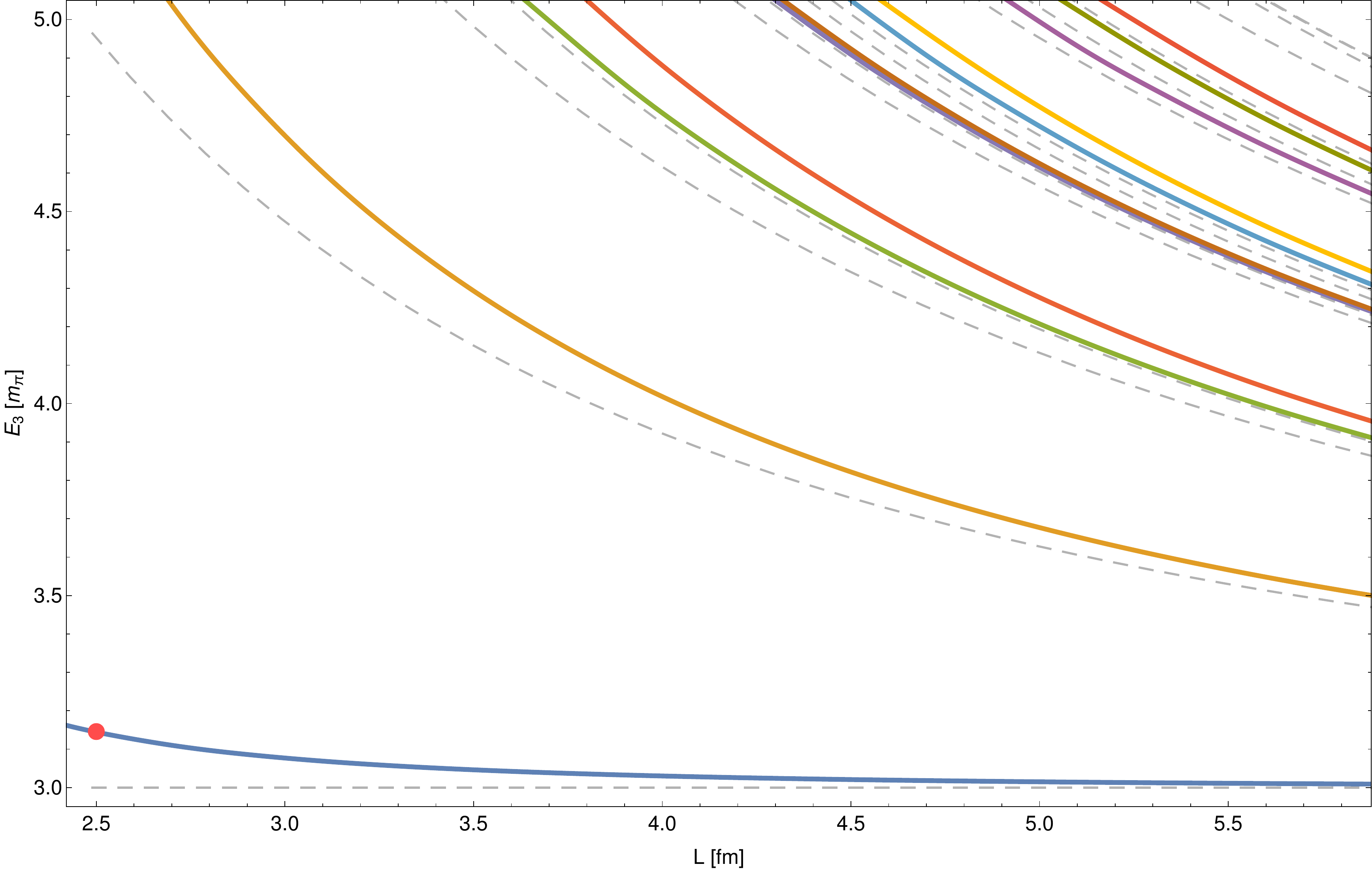}
\caption{
\label{pic:Ldep}
Volume dependence of the three-body energy eigenvalues (colored curves) for $m_\pi=291$~MeV. Dashed gray lines show the positions of the non-interacting levels. Note that the levels 5 and 6, that appear for $L>4.3$~fm, lie almost on top of each other.
}
\end{minipage}
\end{figure*}


\end{widetext}

\end{document}